\documentclass[aps,onecolumn,showpacs,preprintnumbers,amsmath,amssymb] {revtex4}

\newcommand{\p}[1]{\phantom{#1}}
\newcommand{\be}{\begin{equation}}
\newcommand{\ee}{\end{equation}}
\newcommand{\bea}{\begin{eqnarray}}
\newcommand{\eea}{\end{eqnarray}}
\newcommand{\bi}{\begin{itemize}}
\newcommand{\ei}{\end{itemize}}

\newcommand{\data}{---}

\usepackage{graphicx}
\usepackage{amsfonts}
\usepackage{epsfig}

\begin{document}


\title{Holographic renormalization for coincident D$p$-branes}

\author{Toby Wiseman}
\author{Benjamin Withers}
\affiliation{Theoretical Physics, Blackett Laboratory, Imperial
College London, London, SW7 2AZ, UK}

\date{June 2008}

\begin{abstract}
We consider holographic renormalization for the decoupling limit of coincident D$p$-branes. We truncate the theory to the supergravity sector which is homogeneous on the $(8-p)$-sphere and carries only RR electric $(p+2)$-flux, leaving a graviton and two scalar degrees of freedom associated to the dilaton and the  sphere radius. We non-linearly construct the asymptotic graviton and dilaton deformations -- the analog of the Graham-Fefferman expansion for AdS/CFT -- and compute counterterms to give a finite renormalized bulk action and dual one point functions. Restricting to linear deformations we find additional counterterms to include the remaining sphere deformations which strongly deform the asymptotic behaviour. 
\end{abstract}

%
%

\maketitle

%
\section{Introduction}
%

The conjecture of Maldacena's AdS/CFT correspondence \cite{Maldacena:1997re} provides remarkable insight into the structure of non-perturbative string theory and quantum gravity. The holographic dictionary \cite{GKP,Witten:1998qj} between the asymptotic supergravity behaviour of this string theory and the dual CFT, and the subsequent holographic renormalization group developed in 
\cite{Henningson:1998gx,Balasubramanian:1999re,Kraus:1999di,Emparan:1999pm,deHaro:2000xn,Taylor:2001pp,Mueck:2001cy,Bianchi:2001kw,Bianchi:2001de,Berg:2002hy,Skenderis:2002wp,Papadimitriou:2004rz} 
extending previous ideas of Brown and York \cite{Brown:1992br} provide the tools to connect the two sides of the correspondence in detail. Whilst this holographic renormalization group originated for the $AdS_5 \times S^5$ near horizon geometry of decoupled D$3$-branes, it is now taken to apply whenever one considers strings or M-theory on a target spacetime which asymptotically has an AdS factor.

An interesting generalization of AdS/CFT was proposed by Itzhaki et al \cite{Itzhaki:1998dd} where the decoupling limit of coincident D$p$-branes for $p \le 4$ was argued to be dual to 16 supercharge Yang-Mills theory in $(1+p)$-dimensions, the worldvolume theory living on these branes \cite{Witten:1995im}. 
For $p \ne 3$ the theory is no longer conformal, although there does appear to exist a generalization of the conformal symmetry \cite{Jevicki:1998ub}, which is related to the existance of a `dual frame' where the geometry has an AdS factor \cite{Boonstra:1998mp}.  The holographic aspects of the duality have been discussed in the context of D$p$-branes in \cite{Peet:1998wn}. 
The extraction of the dual Yang-Mills thermodynamics from the near extremal behaviour of thermal D$p$-brane black holes \cite{Gibbons:1987ps,Horowitz:1991cd,Garfinkle:1990qj} was performed in \cite{KlebanovTsyetlin}. Several works have addressed the computation of two point functions. In particular for $p=1$ the stress tensor 2 point functions were computed \cite{Hashimoto:1999xu,Antonuccio:1999iz}, and for $p=0$ a full harmonic analysis of the vacuum solution was performed \cite{Sekino:1999av} and all 2 point functions computed. Shortly after it was argued that in analogy to $p=3$, the generalized conformal symmetry actually constrains all 2 point functions to have a simple form \cite{Yoneya:1999bb}.
The map between worldvolume operators and supergravity fields has also been explored in detail\cite{Taylor:1999gq,Sekino:1999av,Taylor:1999pr}. Recently quasinormal mode spectra have been computed \cite{Maeda:2005cr} and shear viscosity extracted from the two point stress energy correlator \cite{Mas:2007ng}. Investigation of the analog of the BMN limit \cite{Berenstein:2002jq} has also been performed \cite{Asano:2003xp,Asano:2004vj}. 

The case of $p \ne 3$ is of considerable interest as it appears that the lower dimensional cases, $p < 3$, are more accessible to direct numerical lattice computation in the Yang-Mills theory due partly to their superrenormalizablity, and partly as fewer dimensions require less lattice points for the same lattice spacing. For $p = 4$ the field theory is non-renormalizable, and hence numerical tests will not be possible. For this reason we will only consider the case $p < 3$ in this paper.
Recent numerical attempts to test the correspondence have been made for $p=1$ by computing the stress tensor two point correlation function \cite{Lunin, Hiller}, and for $p=0$ by computing the theory at finite temperature in order to extract dual black hole entropy \cite{Catterall:2008yz,Catterall:2007fp, Nishimura4, Nishimura16}. For the case $p=0$ there have also been analytic approximation methods used to study this thermal behaviour of theory \cite{Kabat:1999hp,Kabat:2000zv,Kabat:2001ve,Iizuka:2001cw,Oda:2000im}. 

With recent development of exact 16 supercharge supersymmetric lattices \cite{Kaplan3,Catterall:2005fd,Sugino:2006uf,Catterall:2007hk} it is reasonable to hope much numerical progress will be made in the future for the low dimensional cases of the correspondence. This provides a good motivation to consider more carefully the map between supergravity and field theory. Knowing only predictions for two point functions and thermal behaviour provides only limited scope for explicit tests of the correspondence. In principle, if we are able to numerically solve these theories, we would like to perform detailed tests by deforming the field theory with various sources and examining its response, and comparing the prediction from the string theory. It is precisely the technology of holographic renormalization that allows one to make predictions from supergravity for the connection between deformation by sources and the response in expectations values in the field theory.

Whereas the Graham-Fefferman expansion~\cite{Graham} and holographic renormalization for the asymptotic supergravity of the dual string theory has been well developed for the AdS/CFT case $p=3$, it has not been extended to these other non-conformal cases. Holographic renormalization has been successfully extended to a non-AdS/CFT duality only in the case of cascading gauge theories \cite{Klebanov:2000nc,Aharony:2005zr} and little string theory \cite{Marolf:2007ys}. 
We note that previous work has considered the boundary stress tensor specifically for thermal D$p$-brane solutions in \cite{Cai:1999xg} finding consistency with \cite{KlebanovTsyetlin}.

Thus the aim of this paper is to extend the Graham-Fefferman expansion and holographic renormalization to the case of coincident D$p$-branes for $p < 3$. 
We begin the paper in section 2 by discussing the truncation of the string theory dual to $(1+p)$-dimensional maximally supersymmetric Yang-Mills to a particular supergravity sector. In section 3 we use linear theory to identify the physical deformations of the supergravity vacuum solution, and compute their asymptotic form. We study a subclass of these in the full non-linear theory. In section 4 we discuss holographic renormalization for the theory using this asymptotic behaviour of the physical modes. We conclude with a brief discussion in section 5.

%
\section{Near horizon geometry of D$p$-branes}
%

The duality of Itzhaki et al \cite{Itzhaki:1998dd} states that Type IIA/B strings in the near horizon region of $N$ coincident D$p$-branes with $p$ even/odd and $p \le 4$ is dual to 16 supercharge $U(N)$ Yang-Mills in $(1+p)$-dimensions with coupling $g_{YM}$. The string theory may be described by Type IIA/B supergravity provided the string coupling is small and string frame curvatures are small compared to $\alpha' = l_s^2$. 
The bosonic part of this supergravity action in Einstein frame is,
\bea\label{type2action}
S&=&\frac{1}{2\kappa_{10}^2} \int d^{10} x \sqrt{-G} \left(\mathcal{R} -\frac{1}{2}\left(\partial\Phi\right)^2 - g_s^{\frac{(p-3)}{2}} \frac{e^{\frac{(3-p)}{2}\Phi}}{2}\left| F_{p+2}\right|^2\right) 
\eea
where $g_s$ is the string coupling, defined in terms of the dilaton in the asymptotically flat region of the spacetime.
We use coordinates $x^A$, $A = 0,\ldots,9$, the Einstein-frame metric is $G$, we use $\left|F_{p+2}\right|^2 = \frac{1}{\left(p+2\right)!}F_{\sigma_1 \cdots \sigma_{p+2}}F^{\sigma_1 \cdots \sigma_{p+2}}$ and where $F$ is the $(p+2)$-form field strength for the RR-potential that is sourced by the presence of the D$p$-branes and we have neglected the remaining NS-NS and RR form field strengths which may consistently be set to zero. We have taken $2\kappa_{10}^2 = (2\pi)^7 g_s^2 \ell_s^8$. 

The decoupling limit is given by considering fixed $g_{YM}^2 = (2 \pi)^{p-2} g_s \alpha'^{(p-3)/2}$ as $\alpha' \rightarrow 0$. Following Itzhaki et al \cite{Itzhaki:1998dd} finite energy excitations no longer see the asymptotically flat greometry, and instead
 in the supergravity approximation see a string frame metric, dilaton and RR potential,
\bea\label{stringframe}
&& ds^2_{string} =  \alpha' \left( \frac{U^{\frac{7-p}{2}}}{ g_{YM} (d_p N)^{1/2} } \eta_{\mu\nu} \, dx^\mu dx^\nu  + \frac{ g_{YM} (d_p N)^{1/2} }{ U^{(7-p)/2} } \left(dU^2 + U^2 d \Omega_{8-p}^2 \right) \right) \nonumber \\
&& e^\Phi = (2\pi)^{2-p}g_{YM}^2 \left( \frac{U^{7-p}}{g_{YM}^2 d_p N} \right)^{\frac{p-3}{4}}  \nonumber \\
&& A_{01 \ldots p} = \frac{\alpha'^2 U^{7-p}}{g_{YM}^2 d_p N}-1 ,
\eea
where $d_p=2^{7-2p} \pi^{\frac{3(3-p)}{2}}\Gamma\left((7-p)/2\right)$. Here $\mu,\nu = 0,\ldots,p$ are the coordinates transverse to the worldvolume of the D$p$-branes. The radial variable $U$ is interpreted as the energy scale associated to strings stretched between the $N$ D$p$-branes and a probe D$p$-brane at radial position $U$, and hence is thought of as an energy scale in the dual Yang-Mills. We note that as with $p=3$ the interpretation of the energy depends on the precise probe being considered, and supergravity modes infact see an energy scale $E \sim U^{(5-p)/2}/{\sqrt{N g_{YM}^2}}$ \cite{Peet:1998wn}. For $p<3$ the string theory, with effective string coupling $e^\Phi$, is weakly coupled in the UV, $U \rightarrow \infty$. The non-trivial dilaton profile indicates the dual theory is not conformal, which can be simply seen as the Yang-Mills coupling is dimensional. 

The curvature $\mathcal{R}^{string}$ of the string frame metric derives from the size of the $(8-p)$-sphere, and is given in string units by,
\bea
&& \alpha' \mathcal{R}^{string} \sim \frac{U^{3-p}}{N g_{YM}^2}.
\eea
Hence a concern for $p<3$ is that for sufficiently large energies $U$ we see a singular behaviour in the UV boundary region of the vacuum solution. This indicates that the supergravity truncation might break down and we should take into account $\alpha'$ corrections. Since holographic renormalization crucially uses the structure of the supergravity solutions in the near boundary region we might initially ask whether this can be studied self-consistently.

Let us assume we have a fixed characteristic energy scale $U_0$ that is of interest. The dimensionless Yang-Mills effective coupling at this energy scale is $\lambda = N g_{YM}^2 U_0^{p-3}$.
We may then form a dimensionless radial variable $z = U_0/U$. The ultraviolet region of the geometry is then as $z \rightarrow 0$. For $p=3$  the duality is the familiar AdS/CFT correspondence~\cite{Maldacena:1997re}, and the geometry is $AdS_5 \times S^5$ with constant dilaton, with $z$ being the Poincare coordinate, and $z = 0$ being the conformal boundary. 
As for $p = 3$, we see that $z = 0$ is a conformal boundary of the supergravity geometry (independent of frame), and we use the usual 
terminology that it is the \emph{boundary}. For our vacuum solution the string coupling and curvature become,
\bea
&& \alpha' \mathcal{R}^{string} \sim \frac{1}{\lambda} z^{-(3-p)} \nonumber \\
& & e^\Phi \sim \frac{\lambda^{\frac{7-p}{4}}}{N}  z^{(7-p)(3-p)/4}
\eea
and hence for supergravity to be consistent in the bulk for $z \sim O(1)$ we require the strict large $N$, and large $\lambda$ limit. We will be interested in deforming the vacuum solution above for $p<3$ to introduce some characteristic energy scale $U_0$. Having done this we see that the supergravity solution is valid in this limit near the boundary for the range,
\be
\lambda^{-1/(3-p)} < z.
\ee
Thus it is an important point that while the supergravity solution breaks down near the boundary, provided we have large $N$ and large $\lambda$, we may employ the supergravity approximation close enough to the boundary to use the holographic renormalization group.

For the remainder of the paper we will use the Einstein frame metric,
\bea\label{dpgeom}
&& ds^2_{Einstein} = G_{AB} dx^A dx^B = \beta \left(z^{-\frac{(7-p)^2}{8}}\eta_{\mu\nu}dx^\mu dx^\nu+\left(\frac{\lambda d_p}{U_0^2}\right) z^{\frac{-p^2+6p-25}{8}}\left(dz^2 + z^2 d\Omega_{8-p}^2\right) \right) \nonumber \\
&& e^\Phi = \frac{g_s}{ \beta^{\frac{2(3-p)}{(7-p)}}} z^{\frac{(3-p)(7-p)}{4}} \nonumber \\
&& A_{01 \ldots p} = \beta^{\frac{8}{7-p}} \frac{1}{z^{7-p}} -1
\eea
where we have defined the dimensionless combination,
\be
\beta \equiv \left(\frac{\alpha' U_0^2}{\sqrt{\lambda d_p}}\right)^{\frac{(7-p)}{4}},
\ee
and we will use units where $\lambda d_p U_0^{-2}=1$. For convenience we also define,
\be
e^{\phi}  \equiv \frac{\beta^{\frac{2(3-p)}{(7-p)}}}{g_s}  e^{\Phi}. 
\ee
Recall that we have truncated the string theory to the supergravity sector, and have also truncated to only include the RR $(p+2)$-form field strength. For the remainder of the paper we will be concerned with deformations of the above vacuum solution and for convenience we will make the following further consistent truncations. 
Firstly we will only consider deformations that preserve the $SO(9-p)$ isometry of the $(8-p)$-sphere. Hence metric functions will depend only on the worldvolume coordinates $x^\mu$ and the radial coordinate $r$. Secondly we will restrict to deformations of the field strength given by a purely `electric' potential, so that $A_{01\ldots p}$ is a scalar function and all other components of this antisymmetric potential vanish. 

Let us combine the worldvolume coordinates $x^\mu$ and radial $z$ into a set $x^a = \{ x^\mu , z \}$ so $a = 0,\ldots,(p+1)$. Our truncation to homogeneity on the $(8-p)$-sphere allows us to write the 10-d Einstein frame metric as,
\be\label{reductionansatz}
G_{AB} dx^A dx^B = \beta\left(g_{ab}\left(x^\mu,z\right)dx^a dx^b+e^{2 S\left(x^\mu,z\right)}  d\Omega_{8-p}^2\right).
\ee
With our truncation to electric RR fluctuations only, the field strength is constrained by its equations of motion to be,
\be
\left| F_{p+2}\right|^2 = -(7-p)^2 \beta^{\frac{(p^2 -5p+2)}{(7 -p)}}  e^{-(3-p)\phi - 2(8-p) S}
\ee
and we may write a $(p+2)$-dimensional action $S_{trunc}$ that yields the truncated field equations,
\bea\label{strunc}
S_{trunc} & = &  \frac{\beta^4 V_{8-p}}{\kappa_{10}^2} \int d^{p+2}x \sqrt{-g}\, \mathcal{L}_{trunc} \nonumber \\
\mathcal{L}_{trunc} & = & \frac{1}{2} e^{(8-p) S} \Bigg( \mathcal{R}  -\frac{1}{2}\left(\partial \phi\right)^2 + (7-p)(8-p)\left(\partial S\right)^2
 +(7-p)(8-p)e^{-2S}-\frac{(7-p)^2}{2} e^{-\frac{(3-p)}{2}\phi-2(8-p)S}\Bigg) ,
\eea
bearing in mind that this is not simply the result of substituting the flux equation above into the original action \cite{Duff:1989ah}. $V_{8-p} \equiv 2 \left(\pi\right)^{\frac{9-p}{2}} /\Gamma\left(\frac{9-p}{2}\right)$ is the volume of the unit $(8-p)$-sphere. 

We will shortly be concerned with the most general deformations of the above vacuum solution that reside in our consistently truncated sector of the supergravity. In the case that $p=3$ Witten has argued~\cite{Witten:1998qj} that supergravity degrees of freedom with infinite action (non-normalizable deformations) are dual to a source term for a specific operator in the field theory. Finite action (normalizable) deformations for the same degree of freedom are dual to VEV's for that operator. In the case of $p=3$ the program of holographic renormalization, which allows identification of these normalizable and non-normalizable deformations, has now become very well developed~\cite{Papadimitriou:2004rz,Bianchi:2001kw,Bianchi:2001de,deHaro:2000xn,Skenderis:2002wp,Emparan:1999pm,Balasubramanian:1999re,Kraus:1999di,Nojiri:2000kh,Taylor:2001pp,Mueck:2001cy,Berg:2002hy}. The procedure involves constructing a renormalized action which is finite for all deformations. This may be achieved by introducing a geometric cut-off or boundary, and including counterterms that are intrinsic to this boundary. Then removing the boundary yields a finite result. Hence we will later consider the action above, integrated only up to a boundary which we will take to reside at $z = \epsilon$. Calling the \emph{regulated} action $S_{reg}$, we then have,
\bea\label{scutoff}
&&S_{reg}(\epsilon) = \frac{\beta^4 V_{8-p}}{\kappa_{10}^2} \left\{ \int_{z\geq \epsilon} dz\,d^{p+1}x \mathcal{L}_{trunc} - \int_{z=\epsilon}{d^{p}x \sqrt{-\gamma} \mathcal{K}_\gamma} \right\},
\eea
where now we have included the Gibbons-Hawking boundary term where $\gamma$ is the induced metric on the boundary and $\mathcal{K}_\gamma$ is the trace of the second fundamental form.

In considering the deformations of the vacuum solution we will make the following coordinate choice,
\be\label{metricansatz}
g_{ab} dx^a dx^a = \gamma_{\mu\nu}\left(x,z\right)dx^\mu dx^\nu + e^{2 R\left(x,z\right)} dz^2,
\ee
to adapt our coordinate system to the boundary we introduce at $z = \epsilon$. We note that we have not yet locally fixed the gauge. We locally have $(p+2)$ coordinates that we may transform, and having set the off-diagonal metric components to zero constrains $(p+1)$ of these $(p+2)$ local transformations. We might use this remaining freedom to set $R =0$, and hence take a Gaussian normal coordinate system to the boundary, but instead we will leave this freedom explicitly for now, returning to fix it later. In appendix A we give the 10-d field equations  obtained after employing our truncations and coordinate choice above.

%
\section{Near boundary deformations of the vacuum}
%

We begin by considering linear perturbations about the vacuum solution above in order to understand the degrees of freedom present in our truncation, and in addition any residual gauge invariance. From the Lagrangian above we have gravity coupled to 2 scalar fields coming from the dilaton $\phi$ and $(8-p)$-sphere radius $S$. Hence we expect to have degrees of freedom for a graviton in $(p+2)$-dimensions and the 2 scalar degrees of freedom, although we note that these will be combinations of the fields $\phi$, $S$ and the trace of the $(p+2)$-metric. Imagining that we may deform away from our vacuum solution at the boundary and then `integrate in' to the interior we may then consider the data we should specify at the boundary \footnote{We note that of course a hyperbolic system such as the Einstein equations is strictly ill-posed if one specifies data in such a way, although with suitable analytic restrictions a formal power series expansion will exist.}. Introducing a cut-off at $z=\epsilon$ and treating this as a `Cauchy' surface one should specify the values of the scalars on this surface and their momenta, given by their normal derivatives. Let us consider the tensor degree of freedom more carefully. For the graviton one should specify the induced metric on the surface and its momenta. As ADM have shown~\cite{Arnowitt:1960es}, the diffeomorphism invariance implies there are less degrees of freedom than components of this induced metric. The residual coordinate freedom within the slice may be thought of as removing the transverse freedom in the induced metric and the freedom normal to the slice may be thought of as removing the trace. Likewise the constraint equations, the $(zz)$ and $(z\mu)$ components of the Einstein equations, correspondingly constrain the trace and transverse part of the metric momentum. Hence both in the metric and its momentum, the $(p+1)(p+2)/2$ components are reduced to $p(p+1)/2$ actual degrees of freedom.

We will shortly exhibit the linear solutions showing that they capture precisely the expected degrees of freedom. There are two types of deformation; the \emph{graviton/dilaton} and the \emph{sphere}. The graviton/dilaton deformations comprise the graviton and a scalar degree of freedom with the same behaviour near the boundary. Both the field and its momentum must be specified on the cut-off surface for these degrees of freedom. We identify this scalar degree of freedom with the dilaton as the sphere scalar is determined in terms of the dilaton.
The sphere deformation is a scalar degree of freedom, independent of the dilaton, again with value and momentum to specify. As we shall see the sphere deformation behaviour is more pathological near the boundary, and turning it on takes one away from the decoupled near horizon region back to the asymptotically flat D$p$-brane solution.

Having constructed the full linear solution of the near boundary behaviour, we then construct the non-linear extension of the graviton/dilaton mode. It is possible to discuss this mode independently of the sphere mode because the action (\ref{strunc}) admits a further consistent truncation which turns off the sphere deformation:
\be\label{scalartruncation}
S=-\frac{(3-p)}{4(7-p)}\phi.
\ee
Imposing (\ref{scalartruncation}) in the field equations obtained from (\ref{strunc}) one finds that the $S$ and $\phi$ equations of motion become the same, hence this truncation is consistent. Upon imposing this relation at the level of the action (\ref{strunc}), one obtains
\bea\label{onescalar}
S_{graviton/dilaton} & = & \frac{\beta^4 V_{8-p}}{\kappa_{10}^2} \int d^{p+2}x \sqrt{-g}\, \mathcal{L}_{graviton/dilaton} \nonumber \\
\mathcal{L}_{graviton/dilaton} & = & \frac{1}{2} e^{-\frac{(8-p)(3-p)}{4(7-p)}\phi} \Bigg( \mathcal{R}  -\frac{(7-p)^2 p-16}{16(7-p)}\left(\partial \phi\right)^2+\frac{(7-p)(9-p)}{2} e^{\frac{(3-p)}{2(7-p)}\phi}\Bigg),
\eea
which yield the correct truncated equations of motion, whose deformations consist only of the graviton/dilaton mode. Since the sphere mode has a pathological behaviour, as we shall discuss, we do not discuss it non-linearly.

%
\subsection{Linearized analysis of deformations}\label{lineardeformations}
%

Before we proceed to discuss the linear solution we must address the residual gauge invariance. There are two sources of this. Firstly we have not fixed the lapse function $R$ in our metric ansatz. This means that we expect locally to have one function parameterize this freedom and to be undetermined in any solution. Secondly, the metric form is invariant under a coordinate transformation only depending on $x$.

We now use a power series expansion in $z$ to solve the Einstein equations near the boundary $z = 0$ and show that we do indeed capture the 2 scalar and graviton degrees of freedom correctly. The ansatz we take is,
\bea\label{linearexpansions}
\gamma_{ij}&=&z^{\frac{-(7-p)^2}{8}}\left(\eta_{ij}+z^q\sum_{n=0}^{\infty}{\overset{\ell_q,n}{\gamma_{ij}}(x)\; z^{n(5-p)}} \right),\nonumber\\
2 R&=&\frac{-p^2+6p-25}{8}\log{z}- z^q\sum_{n=0}^{\infty}{\overset{\ell_q,n}{R}(x)\; z^{n(5-p)}},\nonumber\\
2 S&=&\frac{-(p-3)^2}{8}\log{z}- z^q\sum_{n=0}^{\infty}{\overset{\ell_q,n}{S}(x)\; z^{n(5-p)}},\nonumber\\
\phi&=&\frac{(7-p)(3-p)}{4}\log{z}+ z^q\sum_{n=0}^{\infty}{\overset{\ell_q,n}{\phi}(x)\; z^{n(5-p)}},
\eea
where we must solve for $q$ from the indicial equations which result from the lowest power of $z$ that arises when expanding the Einstein equations. We have introduced the label $\ell_q$ to indicate the value of $q$ to which the deformation belongs, defined in each case below.

We now state the linear solutions to the Einstein equations. First we consider the leading behaviours allowed by the indicial equations,
\begin{center}
\begin{tabular}{|l |c || c | c| c| c|}
\hline
Mode &  $\ell_q$ & $\overset{\ell_q,0}{\gamma_{ij}}$ &$\overset{\ell_q,0}{R}$ & $\overset{\ell_q,0}{S}$ & $\overset{\ell_q,0}{\phi}$\\
\hline\hline
$q=-(7-p)$ & $\mathbf{-1}$ & $- \frac{(7-p)}{2(3-p)}\eta_{ij}\overset{\mathbf{-1},0}{\phi}$ & $(8-p)\overset{\mathbf{-1},0}{S} + \frac{(7-p)(p+1) }{2 (3-p)}\,\overset{\mathbf{-1},0}{\phi }$ & \data & \data\\
\hline
$q=2(7-p)$ & $\mathbf{2}$ & $- \frac{(7-p)}{2(3-p)}\eta_{ij}\overset{\mathbf{2},0}{\phi}$ & $\frac{2(8-p)(3-p)(5p-33) \overset{\mathbf{2},0}{S} +(7-p)(5p^2-34p+9)\overset{\mathbf{2},0}{\phi}}{2(9-p)(3-p)}$ & \data & \data \\
\hline
$q=0$ &  $\mathbf{0}$ & \data & $\frac{\left(p^2-9 p+18\right) }{2 (p-7)}\,\overset{\mathbf{0},0}{\phi }$ & $\frac{(3-p)}{2(7-p)}\, \overset{\mathbf{0},0}{\phi}$ & \data\\
\hline
$q=(7-p)$ &  $\mathbf{1}$ & constraint (\ref{linearconstraint})  & $\frac{\left(p^2-9 p+18\right) }{2 (p-7)}\,\overset{\mathbf{1},0}{\phi }+\eta^{ij}
   \,\overset{\mathbf{1},0}{\gamma_{ij}}$ & $\frac{(3-p)}{2(7-p)}\, \overset{\mathbf{1},0}{\phi}$ & \data\\
\hline
\end{tabular}
\end{center}
where $\overset{\mathbf{1},0}{\gamma_{ij}}$ is subject to the constraint equation,
\begin{equation}\label{linearconstraint}
4(7-p)\eta^{mn}\partial_m \overset{\mathbf{1},0}{\gamma_{ni}} - 2(5-p)\partial_i \eta^{mn}\overset{\mathbf{1},0}{\gamma_{mn}}+(3-p)(7-p)\partial_i \overset{\mathbf{1},0}{\phi}=0.
\end{equation}
Data which is undetermined in this near boundary analysis is indicated by `\data'. Notice that $\overset{\ell_q,0}{\phi}$ is free in each mode and, as we shall see shortly, parameterizes part of our local residual gauge freedom. Secondly, the subleading terms in each expansion are then determined by the recursion relations for
$n>0$ in terms of the deformations at order $n-1$. The remainder of the local gauge freedom is exhibited in the fact that these recursion relations do not determine any $\overset{\ell_q,n}{\phi}$. The relations we find are,
\bea \label{eq:recursion}
\overset{\ell_q,n}{S} &=&  \frac{(3-p)}{2(7-p)}\overset{\ell_q,n}{\phi}\nonumber\\
&&+\frac{(9-p) \overset{\ell_q,n-1}{ \square_\eta R}+(2 n (p-8) (p-5)+27 p+16 q-2 p (p+q)-89)  \overset{\ell_q,n-1}{\square_\eta S}}{(p-7)^2 (n
   (p-5)-2 p-q+14) (n (p-5)+p-q-7)}\nonumber\\
   &&+\frac{2 (-n (p-5)+p+q-5) \overset{\ell_q,n-1}{\mathcal{R}}}{(p-7)^2 (n
   (p-5)-2 p-q+14) (n (p-5)+p-q-7)},
   \eea
   \bea
\overset{\ell_q,n}{R}&=& \frac{\left((p-3)^2+16 n (p-5)-16 q\right) \overset{\ell_q,n}{\phi}}{2 (p-7) (p-3)}+\frac{1}{3} \Bigg(\frac{(p-9) (p-8)}{(p-7)^3 (-n (p-5)-p+q+7)}\nonumber\\
&&+\frac{(5 p-33) (p-8)}{(p-7)^3 (-n (p-5)+2 p+q-14)}-\frac{6 p}{(p-3)^2 (-n
   (p-5)+p+q-7)}\Bigg)  \overset{\ell_q,n-1}{\square_\eta R}\nonumber\\
   &&+\frac{1}{3} (p-8) \Bigg(\frac{6 p}{(p-3)^2 (-n (p-5)+p+q-7)}+\frac{(2 p-15) (5 p-33)}{(p-7)^3 (n (p-5)-2
   p-q+14)}\nonumber\\
   &&+\frac{(57-4 p) p-201}{(p-7)^3 (n (p-5)+p-q-7)}+\frac{96}{(p-7)^2 (p-3)^2}\Bigg)  \overset{\ell_q,n-1}{\square_\eta S}\nonumber\\
   &&+\frac{2}{3} \Bigg(\frac{2 (p-8) (p-6)}{(p-7)^3 (n
   (p-5)+p-q-7)}-\frac{3 (p-1)}{(p-3)^2 (-n (p-5)+p+q-7)}\nonumber\\
   &&+\frac{(p-8) (5 p-33)}{(p-7)^3 (-n (p-5)+2 p+q-14)}-\frac{48}{(p-7)^2 (p-3)^2}\Bigg)
   \overset{\ell_q,n-1}{\mathcal{R}},
  \eea
  \bea 
    \overset{\ell_q,n}{\gamma_{ij}} &=& -\frac{(7-p)}{2(3-p)}  \overset{\ell_q,n}{\phi}\,\eta_{ij}\nonumber\\
  && +\frac{(p-9)  \overset{\ell_q,n-1}{\square_\eta R} +(8-p) ((2 n-1) (p-5)-2 q)  \overset{\ell_q,n-1}{\square_\eta S} +(2 (n-1) (p-5)-2 q)
   \overset{\ell_q,n-1}{\mathcal{R}}}{(p-3)^2 (n (p-5)-q) (n (p-5)-p-q+7)}\eta_{ij}\nonumber\\
  && +\frac{\partial_i\partial_j \overset{\ell_q,n-1}{R}+(8-p)
   \partial_i\partial_j \overset{\ell_q,n-1}{S}+2 \overset{\ell_q,n-1}{\mathcal{R}_{ij}}}{(n (p-5)-q) (n (p-5)-p-q+7)},
\eea
where $\square_\eta = \eta^{ij}\partial_i \partial_j$. Up to the required order, there is only one divergence in these recursion relations at the following point:
\be\label{divergentpoints}
(p,q,n) = (2,-7+p,5).
\ee
One may worry about the choices of $(1,-7+p,3)$ and $(1,0,3)$ where some of the denominators in the recursion relations vanish, however it is easily verified using the recursion relations for $n<3$ that these cancel amongst themselves. The true divergence found here corresponds to the leading data of the modes $q=-(7-p)$ and $q=2(7-p)$ coinciding at order $z^{10}$. The coincidence of data at a value of $z$ in general means that one should include logarithmic terms in the $z$ expansion at that order. For example, in the case where $p=3$ it is this logarithm which gives rise to the field theory conformal anomaly~\cite{Henningson:1998gx}. In section \ref{linearsphere} we will indeed see that the $p=2$ case will contribute a logarithmic term to the one-point functions.

%
\subsection{A preferred gauge: homogeneous dilaton}
%

As mentioned above the freedom of $\overset{\ell_q,n}{\phi}$ for $n \ge 0$ in each solution above is a result of our local residual gauge freedom due to the unspecified lapse. We now explicitly show this using the coordinate transformation,
\bea
z \to z' &=& z\left(1+  F(x,z)\right),\\
x^{i} \to x^{'i} &=& x^{i}+  \eta^{ij}\partial_j G(x,z),
\eea
where we linearize in the functions $F$ and $G$. These transformations induce an infinitesimal shift in the fields 
\bea
\delta{\phi}(x,z)&=& \frac{(3-p)(7-p)}{4} F,\\
\delta{\gamma_{ij}}&=&z^{-\frac{(7-p)^2}{8}}\left(-\frac{(7-p)^2}{8} \eta_{ij} F + 2 \partial_i\partial_j G\right),\\
\delta{\gamma_{iz}}&=&z^{-\frac{(7-p)^2}{8}}\partial_i\left(\partial_z G + z^{4-p} F\right),\\
\delta{(2R)}&=&-\frac{(3-p)^2}{8}F + 2z\partial_z F,\\
\delta{(2S)}&=&-\frac{(3-p)^2}{8}F.
\eea
In order to preserve our metric ansatz the transformation must have $\delta{g_{iz}}=0$ and hence $
G = \int dz z^{4-p} F$. Then taking $F$ to have the following $z$ expansion,
\bea
F(x,z)&=&z^q\sum_{n=0}^{\infty}{\overset{\ell_q,n}{F}(x)\; z^{n(5-p)}},
\eea
we see the transformation preserves the form of our metric, and also of our power series solution above.
Hence our gauge freedom is given in terms of the functions $\overset{\ell_q,n}{F}(x)$ for $n \ge 0$ which parameterize the local freedom of our unspecified lapse.

We may now consider various gauge choices. One obvious choice is that one may, for any $p$, take the lapse perturbation $\overset{\ell_q,n}{R}$ to vanish, giving the analog of Gaussian normal coordinates.
However, provided $p \ne 3$ we may instead choose to set either $\overset{\ell_q,n}{\phi}$,  $\overset{\ell_q,n}{S}$ or some combination of them to zero by the appropriate choice of $\overset{\ell_q,n}{F}(x)$.

The gauge we choose to work with is that where the dilaton is unperturbed, ie. $\overset{\ell_q,n}{\phi}=0$. We will term this the `homogeneous dilaton' gauge, as the dilaton $\phi$ is then only a function of $z$. Hence we choose our coordinate system such that the relation between our radial coordinate $z$ and the value of the dilaton is fixed. For the cases $p \ne 3$ this is a very physical gauge to choose, as the varying dilaton does pick a preferred radial slicing, and in particular since the dilaton profile is dual to the running of the Yang-Mills coupling in this gauge we have ensured in the dual picture that we are renormalizing our Yang-Mills theory such that the coupling only depends on energy and not position, as is natural from the field theory point of view. While this gauge is possible for a dual with running coupling, and hence non-trivial dilaton profile, in the conformal case $p=3$ since the vacuum has constant dilaton one clearly cannot take this gauge, and it is conventional instead to use the Gaussian-Normal choice $\delta R = 0$, familiar from the Graham-Fefferman expansion.

Another important reason that the homogeneous dilaton gauge is attractive is that from the bulk string theory perspective there are several natural choices of frame - the Einstein frame which we are using, the string frame or the `dual' frame. The metric in each frame differs by a conformal factor which is a power of $e^\phi$. As is usual when considering choices of frame, nothing physical should depend on which choice one makes. An attractive feature of our gauge is that this is manifest rather simply in that under a change of frame, the only change to our near boundary expansion \eqref{linearexpansions} is the leading $z$ behaviour, and the coefficients $\overset{\ell_q,n}{\gamma_{ij}}(x),  \overset{\ell_q,n}{R}(x), \overset{\ell_q,n}{S}(x)$ are invariant, as are their recursion relations \eqref{eq:recursion}.

Note that in addition to the local coordinate freedom above, we may also perform a coordinate transformation $x^{i} \to x^{'i} = x^{i}+ \xi^i(x)$, which only depending on the $x$ coordinates is the remaining global residual freedom.

%
\subsection{The physical modes}\label{physicalmodes}
%

Let us now reconsider the behaviour of our linear deformations using the homogenous dilaton gauge. Now $\overset{\ell_q,n}{\phi} = 0$ for $n \ge 0$. We have,
\begin{center}
\begin{tabular}{|l | c ||c | c| c|| l|}
\hline
Mode & $\ell_q$ & $\overset{\ell_q,0}{\gamma_{ij}}$ & $\overset{\ell_q,0}{R}$ & $\overset{\ell_q,0}{S}$ &  Description\\
\hline\hline
$q=-(7-p)$ & $\mathbf{-1}$ & $0$ & $(8-p)\overset{\mathbf{-1},0}{S}$ & \data & \emph{Sphere non-normalizable}\\
\hline
$q=2(7-p)$ & $\mathbf{2}$ & $0$ & $\frac{(8-p)(5p-33) }{(9-p)}\overset{\mathbf{2},0}{S}$ & \data & \emph{Sphere normalizable}\\
\hline
$q=0$ & $\mathbf{0}$ & \data & $0$ & $0$ & \emph{Graviton/Dilaton non-normalizable}\\
\hline
$q=(7-p)$ & $\mathbf{1}$ & constraint (\ref{linearconstraintfixed}) & $\eta^{ij}\,\overset{\mathbf{1},0}{\gamma_{ij}}$ & $0$ & \emph{Graviton/Dilaton normalizable}\\
\hline
\end{tabular}
\end{center}
where $\overset{\mathbf{1},0}{\gamma_{ij}}$ is subject to the constraint equation,
\begin{equation}\label{linearconstraintfixed}
4(7-p)\eta^{mn}\partial_m \overset{\mathbf{1},0}{\gamma_{ni}} - 2(5-p)\partial_i \eta^{mn}\overset{\mathbf{1},0}{\gamma_{mn}} =0.
\end{equation}

Let us now count our degrees of freedom. We see in our homogenous dilaton gauge the modes $q = -(7-p), + 2(7-p)$ are scalar deformations only involving the $(8-p)$-sphere. Hence these constitute one scalar and its momentum degree of freedom. The remaining non-trivial modes, $q = 0, q = (7-p)$ are deformations only of the metric, leaving the sphere unperturbed. Clearly these include the graviton degrees of freedom we expect to find. Note that the transverse part of $\overset{\mathbf{0},0}{\gamma_{ij}}$ is unphysical due to the global coordinate freedom $x^{i} \to x^{'i} = x^{i}+ \xi^i(x)$, and likewise the momentum $q = (7-p)$ mode has the explicit constraint \eqref{linearconstraintfixed} on the transverse part.
However, we see that in this gauge the trace of $\gamma_{ij}$ is unconstrained for both $q=0, (7-p)$, and hence this trace forms the remaining scalar degree of freedom that we expect to find. The trace of the metric and the scalar degrees of freedom mix in general, and since here $S$ is unperturbed we may attribute this scalar to the dilaton degree of freedom. 

Hence we use the terminology that the $q = -(7-p), + 2(7-p)$ are the \emph{Sphere} deformations, and the $q = 0, q = (7-p)$ are the \emph{Graviton/Dilaton} deformations. We note now, and will discuss in more detail later, that the  $q = -(7-p)$ sphere and $q = 0$ graviton/dilaton modes of deformation give an infinite regulated action as the cut-off surface is removed by taking $\epsilon \rightarrow 0$. Hence we term this \emph{non-normalizable}. Conversely the $q =+ 2(7-p)$ sphere and $q = +(7-p)$ graviton/dilaton modes are of finite regulated action and we term them \emph{normalizable}.

We think of the non-normalizable part of the sphere and metric/dilaton deformations as being `Dirichlet' data that must be fixed at the boundary, and the normalizable part  of the sphere and metric/dilaton deformations are then determined dynamically by the solution of the Einstein equations. In the duality the non-normalizable modes determine the sources for certain operators that deform the field theory action, and the normalizable modes determine the VEVs of these operators.

Taking the limit $p \rightarrow 3$, we indeed recover the expected scalings of the perturbations with the Poincare coordinate $z$. The graviton/dilaton being marginal operators have a non-normalizable $z^0$ and normalizable $z^4$ behaviour. Hence we expect that the graviton/dilaton non-normalizable deformations are dual to sources for the stress-tensor and Lagrangian density (note that the trace of the stress-tensor is simply the Lagrangian density). For $p = 3$ the sphere deformation mode has non-normalizable behaviour $z^{-4}$ corresponding to an irrelevant operator. This mode has been discussed in \cite{Skenderis:2006di,Intriligator:1999ai,Gubser:1998kv,Gubser:1998iu}, and is conjectured to be dual to the leading DBI correction to the open string Yang-Mills description, ie. to an operator $\sim \alpha'^2 F^4$. Being irrelevant, one should only consider such a mode perturbatively. Turning such a deformation on by a finite amount `destroys' the asymptotic behaviour of the solution by reversing the decoupling limit (where the DBI corrections vanish). 
Here for $p \ne 3$ it is natural to assume the same remains true, and indeed we see the linear theory predicts the non-normalizable sphere deformation would indeed strongly deform the asymptotic solution if we were to study it non-linearly.
In the following we shall proceed to only treat the graviton/dilaton deformations non-linearly, and will consider the sphere perturbation linearly.

%
\subsection{Non-linearized analysis of graviton/dilaton deformations}
%

We now perform a nonlinear analysis for the graviton/dilaton deformations, setting to zero the sphere perturbation.  We again compute the graviton/dilaton near boundary behaviour as a power series expansion in $z$ proceding up to $z^{7-p}$ order which will enable us to perform the holographic renormalization in the following section. The computation is performed in the homogeneous dilaton coordinates, and so $\phi$ is undeformed from its form in the vacuum solution.
The power series expansion takes the form analogous to that in the linear theory,
\bea\label{expansionsnonlin}
\gamma_{ij}&=&z^{\frac{-(7-p)^2}{8}}\left( \overset{\mathbf{0},0}{\gamma_{ij}}(x)+\overset{\mathbf{0},1}{\gamma_{ij}}(x)z^{5-p}+\overset{\mathbf{1},0}{\gamma_{ij}}(x)z^{7-p}+\ldots\right),\nonumber\\
2 R&=&\frac{-p^2+6p-25}{8}\log{z}- \left(\overset{\mathbf{0},0}{R}(x)+\overset{\mathbf{0},1}{R}(x)z^{5-p}+\overset{\mathbf{1},0}{R}(x)z^{7-p}+\ldots\right),\nonumber\\
2 S&=&-\frac{(p-3)^2}{8}\log{z}-  \left(\overset{\mathbf{0},0}{S}(x) +\overset{\mathbf{0},1}{S}(x) z^{5-p}+\overset{\mathbf{1},0}{S}(x) z^{7-p}+\ldots\right),\nonumber\\
\phi&=&\frac{(7-p)(3-p)}{4}\log{z},
\eea
up to the order of interest. The non-linear generalization of the $q = 0, (7-p)$ metric/dilaton modes are then given by,
\begin{center}
\begin{tabular}{|l |c| c|| c | c| c|}
\hline
Mode & $\ell_q$ & $n$ & $\overset{\ell_q,n}{\gamma_{ij}}$ & $\overset{\ell_q,n}{R}$ & $\overset{\ell_q,n}{S}$\\
\hline\hline
$q=0(7-p)$ & $\mathbf{0}$ &  $0$ & \data & $0$ & $0$\\
\hline
$q=0(7-p)$ & $\mathbf{0}$ &  $1$ & $\frac{1}{(p-5)}\overset{\mathbf{0},0}{\mathcal{R}_{ij}}$ & $\frac{1}{ (p-9)} \overset{\mathbf{0},0}{\mathcal{R}}$ & $0$\\
\hline
$q=1(7-p)$ & $\mathbf{1}$ & $0$ & constraint (\ref{nonlinearconstraintfixed}) & $\overset{\mathbf{0},0}{\gamma^{ij}}
   \,\overset{\mathbf{1},0}{\gamma_{ij}}$ & $0$\\
\hline
\end{tabular}
\be\label{nonlinearsols}
\ee
\end{center}
where $\overset{\mathbf{0},0}{\mathcal{R}}$ and $\overset{\mathbf{0},0}{\mathcal{R}_{ij}}$ are the Ricci scalar and tensor of the metric $\overset{\mathbf{0},0}{\gamma_{ij}}$, and where $\overset{\mathbf{1},0}{\gamma_{ij}}$ is subject to the constraint equation,
\begin{equation}\label{nonlinearconstraintfixed}
4(7-p)\overset{\mathbf{0},0}{\nabla^j} \overset{\mathbf{1},0}{\gamma_{ji}}-2(5-p)\partial_i \overset{\mathbf{0},0}{\gamma^{mn}}\overset{\mathbf{1},0}{\gamma_{mn}}=0.
\end{equation}

%
\section{Holographic renormalization}
%

In the previous section we have computed the near boundary behaviour for deformations of the vacuum. In this section we proceed to compute the regulated action evaluated on-shell on our near boundary solutions above. We denote the regulated action evaluated on-shell as $S^{on-shell}_{reg}(\epsilon)$. The divergences in this when one removes the regulator are, by definition, due to non-normalizable deformations.
These divergences arise in our deformations from only a few leading terms in the expansions of the non-normalizable deformations. Hence only knowledge of the first few terms is typically required to compute them.
Following \cite{Papadimitriou:2004rz,Bianchi:2001kw,Bianchi:2001de,deHaro:2000xn,Skenderis:2002wp} we renormalize the on-shell action by introducing counterterms on the regulator boundary, $S_{ct}(\epsilon)$, yielding a new renormalized action $S_{renorm} = \lim_{\epsilon \rightarrow 0} \left( S^{on-shell}_{reg}(\epsilon) - S_{ct}(\epsilon) \right)$ which is finite when the regulator is removed, and is a function of the data in our on-shell deformation.

Following the usual prescription the non-normalizable modes are dual to sources for certain operators in the dual Yang-Mills. Then evaluating the renormalized action on our deformations, and varying with respect to the data of the non-normalizable modes allows us to compute the dual VEVs for these operators. Whilst evaluating the counterterm variations is simple as contributions only come from the boundary itself, $S^{on-shell}_{reg}$ involves an integral over all the radial coordinate, and therefore whilst computing divergences only requires a few leading terms of the $z$ expansions  to be known, naively computing variations with respect to the solution data requires knowledge to all orders in our $z$ expansions. However part of the elegance of holographic renormalization is that this is not the case.
In fact varying $S^{on-shell}_{reg}$ one may then perform an integration by parts to yield a contribution on the boundary, and a bulk contribution that vanishes by virtue of the equations of motion. One then finds,
\bea\label{svar}
\delta S^{on-shell}_{reg}(\epsilon) &=& \frac{\beta^4 V_{8-p}}{2\kappa_{10}^2}\int_{z=\epsilon} d^{p+1}x \sqrt{-\gamma} e^{(8-p)S} \Bigg(\left[ e^{-R}\partial_z \phi\right] \delta \phi - \left[2(8-p)\left(\mathcal{K}+(7-p)e^{-R}\partial_z S\right)\right]\delta S\nonumber\\
&& -\left[\mathcal{K}_{ij} - \gamma_{ij} \mathcal{K} - \gamma_{ij}(8-p)e^{-R}\partial_z S\right]\delta \gamma^{ij}\Bigg),
\eea
where $\mathcal{K}_{ij}\equiv \frac{e^{-R}}{2}\partial_z \gamma_{ij}$, $\mathcal{K}\equiv tr\left(\mathcal{K}\right)$.
Using this, variations of $S^{on-shell}_{reg}$ evaluated for our power series solutions may be conveniently computed using only the leading terms of the expansion, with higher terms yielding no contribution in the $\epsilon \rightarrow 0$ limit. Note that provided the renormalized action $S_{renorm}$ is indeed finite for all data of our near boundary solutions, then variations of it with respect to these data will automatically be finite.

We have defined our regulator boundary to be at coordinate location $z = \epsilon$. However depending on our coordinate choice, this may be at different physical locations.
The obvious choice is to take the regulator to be on constant dilaton surfaces. 
We have discussed that for $p \ne 3$ a physical choice of radial coordinate $z$ is precisely that picked out by constant $\phi$ surfaces. Hence using this coordinate system our $z = \epsilon$ boundary is indeed a constant dilaton surface. We will use the homogeneous dilaton (HD) coordinate system discussed above, and the counter terms we find will be those particular to that gauge. Hence we will call our counter term action $S_{ct}^{HD}$ to emphasize that one must use the homogeneous dilaton coordinates when computing the on-shell renormalized action. It would be a straightforward exercise to generalize the action to not be restricted to this gauge, but we note that one should also include the off-diagonal components of \eqref{metricansatz} in order to be fully general and for simplicitly we do not consider this further here.

We begin by computing the renormalized action for non-linear deformations of the graviton/dilaton. 
We then proceed to extend this in the linear theory to include counterterms for the sphere deformations. 

%
\subsection{Holographic renormalization of graviton/dilaton deformations}
%

The action evaluated on-shell for graviton/dilaton deformations takes the following form in its $\epsilon$ expansion,
\be\label{sregonshellnonlin}
S_{reg}(\epsilon) \sim \frac{\beta^4 V_{8-p}}{2\kappa_{10}^2}\int_{z=\epsilon} d^{p+1}x \sqrt{-\overset{\mathbf{0},0}{\gamma}} \left(\frac{\overset{\mathbf{0},0}{f(x)}}{\epsilon^{7-p}}+\frac{\overset{\mathbf{0},1}{f(x)}}{\epsilon^2}+\mathcal{O}\left(\epsilon^0\right)\right),
\ee
where $\overset{\ell_q,n}{f(x)}$ denotes some function of $q$-mode deformations up to order $n$. Thus there are two terms which diverge as the cutoff $\epsilon$ is removed and we expect to have to add counterterms with no derivatives to cancel the leading divergence, and two derivatives to cancel the subleading one.
We will employ our physically motivated homogeneous dilaton gauge now. Hence on the regulating surface the value of the dilaton will be constant, and simply related to $\epsilon$ as for the vacuum solution.
We therefore use the ansatz,
\bea\label{ctnonlin}
&&S^{HD}_{ct}(\epsilon) = \frac{\beta^4 V_{8-p}}{2\kappa_{10}^2}\int_{z=\epsilon} d^{p+1}x \sqrt{\gamma}\left( A e^{c_1\phi + c_2 S} +B e^{c_3\phi + c_4 S}+ C e^{c_5\phi + c_6 S} \mathcal{R}_\gamma \right),
\eea
where $\mathcal{R}_\gamma$ is the Ricci scalar of the induced metric at the regulator boundary $\gamma_{ij}$. We have not included boundary derivatives of $\phi$ in the action as $\phi$ is constant on the regulator surface. In addition since $S$ is also homogeneous on the boundary surface for the deformations in question we also have not included its derivatives in the action. 

Substituting our deformations (\ref{expansionsnonlin},\ref{nonlinearsols}) into the regulated action and comparing with the counterterm action, we find that the renormalized action is finite provided we choose,
\be
A+B = 9-p \qquad C = \frac{1}{2}  \nonumber 
\ee
\be
c_1=\frac{(3-p)}{4(7-p)}\left(c_2+p-7\right),\qquad c_3=\frac{(3-p)}{4(7-p)}\left(c_4+p-7\right), \qquad c_5=\frac{(3-p)}{4(7-p)}\left(c_6+p-9\right).
\ee
We may now vary the renormalized action with respect to the non-normalizable data, $\overset{\mathbf{0},0}{\gamma^{ij}}$, in our deformation. 
Following from $p=3$ we interpret $\overset{\mathbf{0},0}{\gamma^{ij}}$ as the metric for the dual Yang-Mills theory, and hence the source for the Yang-Mills stress tensor. Then the expectation value of this stress tensor, $\left< T_{ij} \right>$, is given by this variation,
\be\label{efvariationnonlin}
\delta S_{ren}=\underset{\epsilon\to 0}{\ell im} \left(\delta S_{reg}(\epsilon)-\delta S^{HD}_{ct}(\epsilon)\right)=\int d^{p+1}x \left(- \frac{1}{2} \sqrt{-\overset{\mathbf{0},0}{\gamma}} \, \delta \overset{\mathbf{0},0}{\gamma^{ij}} \, \left< T_{ij} \right>   \right),
\ee
and explicit computation gives,
\bea\label{stressenergynonlin}
\left< T_{ij} \right> &=& \frac{\beta^4 V_{8-p}}{4\,\kappa_{10}^2} \left(+2(7-p)\overset{\mathbf{1},0}{\gamma_{ij}}-(5-p)\overset{\mathbf{0},0}{\gamma^{mn}}\overset{\mathbf{1},0}{\gamma_{mn}}\overset{\mathbf{0},0}{\gamma_{ij}}\right) .
\eea
As a consequence of the $(iz)$ Einstein constraint equation we find that the dual field theory stress tensor is conserved in the metric $\overset{\mathbf{0},0}{\gamma_{ij}}$, so that,
\be\label{conservationnonlin}
\overset{\mathbf{0},0}{\nabla^i}  \left< T_{ij} \right> =0 .
\ee
Whereas in the $p = 3$ case the trace of the stress tensor is determined simply by the conformal anomaly and not dynamically, in our case here, $p \ne 3$, the trace is dynamical and is governed by the trace of the metric deformation which in our gauge is the dilaton scalar degree of freedom.
The trace of the stress-tensor is simply proportional to the Lagrangian density of the Yang-Mills theory, and hence we see that the dual operator to the trace of the metric is this Lagrangian density as we should expect for the dual to the dilaton degree of freedom which directly determines the Yang-Mills coupling.

%
\subsection{Example: Thermal D$p$-brane}\label{thermaldp}
%

As an example we consider the particular deformation due to turning on finite temperature. Then the solution of interest is the decoupled limit of the thermal D$p$-brane solution. 
This is as for the vacuum solution given above, except that the Einstein-frame metric is deformed as \cite{Gibbons:1987ps,Horowitz:1991cd,Garfinkle:1990qj}, 
\bea\label{thermaldpbrane}
ds^2 &=& \tilde{z}^{-\frac{(7-p)^2}{8}} \left(-1+\left(\frac{r_0}{r_-}\right)^{7-p} \tilde{z}^{7-p} \right)dt^2 + \tilde{z}^{-\frac{(7-p)^2}{8}} \delta_{ij} dx^i dx^j\nonumber\\
&+&  r_-^2\tilde{z}^{\frac{-p^2+6p-25}{8}}\left(1-\left(\frac{r_0}{r_-}\right)^{7-p} \tilde{z}^{7-p} \right)^{-1}d\tilde{z}^2 +  r_-^2 \tilde{z}^{-\frac{(3-p)^2}{8}}d\Omega_{8-p}^2,
\eea
where $r_0^{7-p}\equiv r_+^{7-p}-r_-^{7-p}$, and $r_-^{7-p}= d_p \lambda \alpha'^{5-p} U_0^{3-p}$. Conveniently this solution is already in our homogeneous dilaton coordinate system. Performing a simple rescaling of $\tilde{z}$,
\be
\tilde{z} = \beta^{-\frac{8}{(7-p)^2}} z,
\ee
we obtain the metric in a form compatable with our ansatz,
\bea\label{thermaldpbrane2}
ds^2 &=& \beta \Bigg\{z^{-\frac{(7-p)^2}{8}} \left(-1+\left(\frac{r_0}{r_-}\right)^{7-p} \beta^{-\frac{8}{(7-p)}} z^{7-p} \right)dt^2 +  z^{-\frac{(7-p)^2}{8}} \delta_{ij} dx^i dx^j\nonumber\\
&+&  z^{\frac{-p^2+6p-25}{8}}\left(1-\left(\frac{r_0}{r_-}\right)^{7-p} \beta^{-\frac{8}{(7-p)}} z^{7-p} \right)^{-1}dz^2 +   z^{-\frac{(3-p)^2}{8}}d\Omega_{8-p}^2\Bigg\},
\eea
where we have chosen units so $d_p \lambda U_0^{-2} = 1$. The deformation is of our graviton/dilaton type. There is no non-normalizable part as $\overset{\mathbf{0},0}{\gamma_{ij}}$ = 0, and only the normalizable data $\overset{\mathbf{1},0}{\gamma_{tt}}$ is non-vanishing. Calculating the expectation value of the field theory stress-energy tensor from the obtained relations (\ref{stressenergynonlin}), and re-inserting appropriate powers of $d_p \lambda U_0^{-2}$ on dimensional grounds we find
\bea
\left<\rho\right>=\left<T_{tt}\right>  = \frac{V_{8-p}}{\kappa_{10}^2}\,\frac{(9-p)}{4}r_0^{7-p}\\
\left<\mathcal{P}\right>=\left<T_{xx}\right>  =\frac{ V_{8-p}}{\kappa_{10}^2}\,\frac{(5-p)}{4}r_0^{7-p},
\eea
where $xx$ denotes any of the spatial worldvolume directions, and $\left<\rho\right>,\left<\mathcal{P}\right>$ are the expectation values of the energy density and pressure respectively. 

In the paper \cite{Cai:1999xg} the field theory stress tensor for this deformation was computed by choosing a frame where the vacuum solution has an $AdS_{p+2}$ factor - the `dual-frame' -  and then using the holographic stress tensor for the $AdS$ part of the metric evaluated on this solution. The result is the same. Furthermore in that reference the energy density and pressure were computed using black hole thermodynamics for the asymptotically flat D$p$-brane solution. Subsequently taking the decoupling limit of these thermodynamic expressions gave rise to the same energy density and pressure, showing consistency with the holographic approach.

%
\subsection{Holographic renormalization of linearized sphere deformations}\label{linearsphere}
%

In this section we extend the holographic renormalization analysis to include also the remaining sphere deformation. However, since the non-normalizable part of this strongly changes the asymptotic behaviour of the solution, we only consider this mode and its normalizable component linearly. We use the linear deformations of section \ref{physicalmodes} and extend the counterterm action found above for the graviton/dilaton deformation to this case.
Since $S$ is now not homogeneous when the sphere deformations are turned on we may add to our graviton/dilaton counterterm action (\ref{ctnonlin}) kinetic terms involving $S$ on the boundary. Hence we use the counterterm action,
\bea\label{ctlinear}
S^{HD}_{ct}(\epsilon) &=& \frac{\beta^4 V_{8-p}}{2\kappa_{10}^2}\int d^{p+1}x \sqrt{\gamma}\Bigg( \left[A e^{c_1\phi + c_2 S}+B e^{c_3\phi + c_4 S}+C e^{c_5\phi + c_6 S} \mathcal{R}_\gamma\right]\nonumber\\
&&+\left[D e^{c_7\phi+c_8 S} \square_\gamma S+E e^{c_9\phi+c_{10} S} \square^2_\gamma S+F e^{c_{11}\phi+c_{12}S} \square^3_\gamma S+G e^{c_{13}\phi+c_{14}S} \square^4_\gamma S\right]\Bigg)_{z=\epsilon},
\eea
where the terms in the first bracket are as for the graviton/dilaton calculation above. By expanding the first two terms in this bracket to quadratic order in the deformations one can see that there are only three independent terms, thus one expects a single ambiguity here. For the second bracket only the combinations $D c_8, E c_{10}, F c_{12}$ and $G c_{14}$ are independent parameters in the linear theory, as can be seen by integrating by parts. 

In our linear solution the non-normalizable components are determined by the data $\overset{\mathbf{0},0}{\gamma_{ij}}$ and $\overset{\mathbf{-1},0}{S}$, and give rise to the divergences in the regulated action. To determine the remaining counterterm parameters we look at variations with respect to this data, and require that the divergences cancel. This is equivalent to requiring cancellation of divergences in the action to quadratic order in the deformation, provided the undeformed action is regular. We find that the remaining counterterm parameters are determined as,
\be
\begin{array}{ll}
  A = \frac{2 (p-8) (p-7)^3}{(p-9) c_2^2+4 (p-8) (p-7) c_2+2 (p-8) (p-7)^2} , & D\, c_8= \frac{3p-23}{2},\nonumber\\\\
  E\, c_{10}= \frac{(7-p)^2}{4 (11-p) (9-p) (8-p)}, & F\, c_{12}=\frac{(7-p)^2}{24 (8-p)^2 \left(p^2-20 p+99\right)},\nonumber\\\\
  G\, c_{14}=-\frac{(7-p)^2 (2 p-25)}{48 (11-p)^2 (9-p) (8-p)^3 (1+p)}, & c_4=\frac{2 (8-p) (7-p) (c_2+p-7)}{c_2 (9-p)-2 (8-p) (7-p)}, \nonumber\\\\
  c_6=\frac{4 (8-p)}{(9-p)}, & c_{2k+1}=\frac{(3-p)}{4 (7-p)}(c_{2k+2}+p-3-2k) \nonumber\\
\end{array}
\ee
for $k=3,4,5,6$. Indeed for this set of counterterms the undeformed action is finite, and hence the deformed action is finite too, up to quadratic order in the deformations. Thus we find no ambiguity in the counterterm ansatz for the linear theory.

As discussed above, the sphere non-normalizable mode is dual to a source for a higher-dimension correction $\mathcal{O}_S$ to the Yang-Mills action, thought to be $\sim \alpha'^2 F^4$. Varying the renormalized action with respect to the non-normalizable data $\overset{\mathbf{-1},0}{S}$, we may then obtain the VEV of this operator $\left< \mathcal{O}_S \right>$ which is determined by the normalizable part of the sphere deformation, 
\be\label{efvariation}
\delta S_{ren}=\underset{\epsilon\to 0}{\ell im} \left(\delta S_{reg}(\epsilon)-\delta S_{ct}(\epsilon)\right)=\int d^{p+1}x \left(- \frac{\left<T_{ij}\right>}{2}  \delta \overset{\mathbf{0},0}{\gamma^{ij}}-\frac{\left<\mathcal{O}_S\right>}{2} \overset{\mathbf{-1},0}{\delta S}\right) .
\ee
Computing the variations explicitly, we find
\bea
\left<T_{ij}\right>&=& \frac{\beta^4 V_{8-p}}{4\,\kappa_{10}^2}\left(2(7-p)\overset{\mathbf{1},0}{\gamma_{ij}}-(5-p)\eta^{mn}\overset{\mathbf{1},0}{\gamma_{mn}}\eta_{ij}\right)\label{linearstressenergy}\\
\left<\mathcal{O}_S\right>&=&-\frac{\beta^4 V_{8-p}}{\kappa_{10}^2}\left(\frac{3(8-p)(7-p)^3}{(9-p)}\right)\overset{\mathbf{2},0}{S}\label{linearsvar},
\eea
where the stress tensor is simply the linearized version of our expression above (\ref{stressenergynonlin}). 

For D2-branes there is an additional divergent contribution to the action, proportional to $\overset{\mathbf{-1},0}{\square^5_\eta S} \log{z}$. Obtaining the precise coefficient for this term requires extending our recursion relations to include logarithms at the order $n=5$. This is of course coincident with a divergence of the recursion relations (\ref{divergentpoints}). In order to have a finite action and one point function a new logarithmically divergent counterterm would have to be added to the action, as for the conformal anomaly for $p=3$. 

The log divergence occurs for $p=2$ and depends only on the non-normalizable data $\overset{\mathbf{-1},0}{S}$. Therefore a trick is to deform $p=2$ to $p = 2-|\delta|$, $|\delta| << 1$, where our renormalized action is now finite, as is the corresponding one-point function for $\mathcal{O}_S$ obtained by variation. From this we know that even for $p=2$, the expression for the one point function \eqref{linearsvar} is the correct one provided we added the appropriate log divergent counterterm to the action. 

However, as we have discussed, the non-normalizable sphere deformation appears to only be treatable perturbatively ie. the source in the dual Yang-Mills can't be turned on. Whilst we have to consider this non-normalizable component to use the holographic technology to compute the one-point function for $\mathcal{O}_S$, in practice we would be interested in vacuum deformations that do not have the non-normalizable sphere component in, and hence we should not encounter this log divergence when evaluating the action on-shell.

%
\section{Discussion}
%

We have discussed the holographic renormalization for coincident D$p$-branes, in part motivated by recent numerical lattice attempts to study the strongly coupled Yang-Mills description directly. In the future one would hope that detailed comparisons with the dual closed string theory at least in the supergravity limit could be made, using the holographic dictionary which we are considering here. As we have seen the structure of the analysis is similar to that of the $p=3$ AdS/CFT case, the principle difference being that we have used a physical gauge derived from the running of the dilaton which is not possible for $p=3$. 

In principle holographic renormalization provides a very powerful predictive tool for the dual Yang-Mills theory, relating sources explicitly to specific vacuum expectation values. However in order to use this predictive power bulk solutions must be found and cast in our generalized Graham-Fefferman form in order to read off these predictive relations. The key point is that consideration of the bulk solutions requires imposing physical boundary conditions in the IR of the geometry, as well as fixing the non-normalizable sources in the UV. For $p=0$ as noted earlier a full linear analysis has been performed \cite{Sekino:1999av}, and this might be extended to general $p$ and used to give predictions for the behaviour of one-point functions using our holographic renormalization group. However a linear analysis will not allow one to compute the response of the theory to sources, which manifestly requires non-linear interaction. Progress in developing techniques to solve the non-linear bulk Einstein equations has developed on various fronts, in particular the derivative expansions for hydrodynamics \cite{Bhattacharyya:2008jc} which has recently been generalized to include non-normalizable sources \cite{Bhattacharyya:2008ji}, and also analytic and numerical understanding of non-trivial relevant supergravity solutions \cite{Gibbons:2004js,Harmark:2004ws,Aharony:2004ig,Aharony:2005bm}. Development of the hydrodynamic derivative expansion is so far for AdS/CFT only, but it is clear that it may be extended to the D$p$-brane cases for $p<3$, although being manifestly Lorentzian it is unlikely it will allow lattice tests of the correspondence. It will be interesting in future to see if any of these techniques may applied to find non-trivial bulk solutions,  for various $p$, deformed by boundary sources which are of interest as probes of the correspondence that might be tested on the lattice via the holographic renormalization group.

We have considered deformations of the $(8-p)$-sphere, which we believe from previous arguments made for $p=3$ are dual to DBI corrections to the Yang-Mills action $\mathcal{O}_S \sim \alpha'^2 F^4$.  It is interesting that at the non-linear level in the supergravity these deformations decouple from the graviton/dilaton deformations and can hence be `turned off'. In the context of Yang-Mills lattice simulations, DBI corrections will not be included, and hence one is unlikely to wish to consider the non-normalizable part of this sphere deformation.  
One might naively conclude that the normalizable part of the sphere deformation is also uninteresting, as without  its non-normalizable source turned on it will be absent and hence can only have a trivial one-point function for $\mathcal{O}_S$.
Whilst it is true that without its source it is consistent to set it to zero, this need not be the case.
For example, in the Lorentzian context  one might consider the evolution of initial data in the supergravity that has a normalizable sphere deformation component. More likely to be of interest for numerical simulations of Yang-Mills one might be interested in testing multi-point functions that involve two or more $\mathcal{O}_S$ operators.

Interestingly there seems to be no obstacle in applying the results obtained here to the D-instanton, $p=-1$, for which the worldvolume theory is the IKKT model~\cite{Ishibashi:1996xs}. There are of course no worldvolume directions, and so the only physical deformation would be that of the sphere. Due to its lack of spatial extent, the IKKT model may be the easiest setting in which to numerically simulate a theory with string dual. Developing a better understanding of the DBI corrections to this model, and effect on the supergravity solution of deforming non-linearly by the sphere non-normalizable mode might be a promising direction for new tests. Numerical work has been conducted on this model \cite{Ambjorn}, although so far only employing a loop approximation.

It has been a deliberate choice in this work not to use the `dual frame' discussed in \cite{Boonstra:1998mp}. In this frame, an AdS factor does emerge in the geometry, and in particular the boundary stress tensor can be extracted using similar methods to the usual AdS/CFT case as was done for the specific thermal deformation \eqref{thermaldpbrane} above in \cite{Cai:1999xg}. The reason we have not used the dual frame is that whilst the calculations may appear closer to what one does in the AdS/CFT case, they are no simpler and we feel little intuition is gained this way. The dual theory is not a conformal theory and the running of the dilaton is an important aspect of this and allows us to employ the useful homogenous dilaton gauge. We believe the simplest conceptual way to apply the holographic renormalization is as we have done here, simply to consider physical deformations, identify those of divergent and finite bare action, and construct local counterterms for the bulk regulated action to render it finite when evaluated on-shell.

There are various other interesting directions for future work. We have truncated to the supergravity deformations which are homogeneous on the $(8-p)$-sphere, and also to only electric fluctuations of the $(2+p)$-form RR field strength. Including non-linear deformations of the remaining form fields would be interesting. Likewise, considering the modes of deformation that are inhomogeneous on the sphere would be interesting and could presumably be done using the methods of `Kaluza-Klein' holography discussed for AdS/CFT in \cite{Skenderis:2006uy}. In addition, a consideration of the Coulomb branch of the Yang-Mills theory via the decoupling limit of multi-centered D$p$-brane solutions and comparison to the gauge theory would likely follow along the lines of \cite{Skenderis:2006di} in the AdS/CFT case.

%
\section*{Acknowledgements}
%

We would like to thank Simon Catterall and Shiraz Minwalla for useful discussions. TW is partly supported by a STFC advanced fellowship and Halliday award. BW is supported by an STFC studentship.

\appendix

%
\section{Field Equations}\label{reduction}
%

To obtain the field equations associated with the metric ansatz (\ref{reductionansatz},\ref{metricansatz}) we perform a reduction of the terms in the field equations directly. The full, unreduced field equations for our action (\ref{type2action}) are, for the dilaton,
\be\label{fe1}
\Box\Phi=\left(\frac{3-p}{4}\right) g_s^{\frac{(p-3)}{2}}e^{\frac{3-p}{2}\Phi}\left| F_{p+2} \right| ^2,
\ee
and for the RR form,
\be\label{fe2}
\nabla_\mu \left(g_s^{\frac{(p-3)}{2}} e^{\frac{3-p}{2}\Phi} F^{\mu\sigma_1 \cdots \sigma_{p+1}}\right)=0,
\ee
and the Einstein equations
\be\label{fe3}
\mathcal{R}_{\mu\nu}-\frac{1}{2}G_{\mu\nu}\mathcal{R}=\frac{1}{2}\partial_\mu \Phi \partial_\nu \Phi - \frac{1}{4} G_{\mu\nu} \left(\partial \Phi\right)^2 + \frac{1}{2}\frac{g_s^{\frac{(p-3)}{2}} e^{\frac{3-p}{2}\Phi}}{\left(p+1\right)!}F_{\mu\sigma_1 \cdots \sigma_{p+1}}F_{\nu}^{\p{\nu}\sigma_1 \cdots \sigma_{p+1}}-\frac{1}{4} G_{\mu\nu} g_s^{\frac{(p-3)}{2}} e^{\frac{3-p}{2}\Phi}\left| F_{p+2} \right| ^2.
\ee
Upon inserting our metric ansatz (\ref{reductionansatz},\ref{metricansatz}) and using Cartan's structural equations to obtain the curvature terms, we obtain field equations in terms of the `reduced' quantities. That is, covariance is with respect to the $p+1$-dimensional brane-metric $\gamma_{ij}$, and the metric functions $R$ and $S$ are scalar degrees of freedom.
Primes indicate $z$ derivatives, $\mathcal{K}_{ij}\equiv \frac{e^{-R}}{2}\partial_z \gamma_{ij}$, $\mathcal{K}\equiv tr\left(\mathcal{K}_{ij}\right)$ and $i,j=0,\ldots,p$. $\mathcal{R}$ is the Ricci scalar of $\gamma_{ij}$.\\\\
\underline{Einstein- $ij$:}
\bea
\mathcal{R}_j^i-\frac{1}{2}\delta^i_j\mathcal{R}&=&\left(\nabla_j \partial^i R-\delta _j^i \square R \right)+(8-p)\left(\nabla_j \partial^i S-\delta_j^i \square S \right)\nonumber\\
&&+ \left(\partial^i R\partial_j R- \delta _j^i\left(\partial R\right)^2\right)+(8-p)\left( \partial^i S\partial_j S-\frac{1}{2}(9-p) \delta_j^i \left(\partial S\right)^2\right) \nonumber\\
&&-(8-p)\delta _j^i\left(\partial R \cdot \partial S\right) +\frac{1}{2} \left(\partial^i \phi\partial_j \phi-\frac{1}{2} \delta _j^i \left(\partial \phi\right)^2 \right)\nonumber\\
&&+e^{-2R} \left(-\frac{1}{2} (8-p) (9-p)  S'^2 \delta
   _j^i-\frac{1}{4} \phi'^2 \delta _j^i+ (8-p)  R' S'\delta _j^i-(8-p) S'' \delta _j^i\right)\nonumber\\
&&+e^{-R}\left((8-p)\left(\mathcal{K}^i_j-\delta^i_j\mathcal{K}\right)S'+\left(\mathcal{K}^{i}_j\right)'-\delta_j^i\mathcal{K}'\right)+\left(\mathcal{K}^i_j\mathcal{K}-\frac{1}{2}\delta^i_j\left(\mathcal{K}^2+\mathcal{K}\cdot\mathcal{K}\right)\right)\nonumber\\
&&+\frac{(7-p)}{2}\delta _j^i\left( e^{-2 S} (8-p) -\frac{(7-p)}{2} e^{-\frac{(3-p)}{2}  \phi-2 (8-p) S} \right).
\eea
\underline{Einstein- $z,i$:}
\bea
\nabla^j\left(\mathcal{K}^i_j-\delta^i_j \mathcal{K}\right)&=&-(8-p)\mathcal{K}^i_j\partial^j S- e^{-R}\left((8-p)\left(S'\partial^i R-S' \partial^i S - \partial^i S'\right)-\frac{1}{2}\phi' \partial^i \phi\right).
\eea
\underline{Einstein- $zz$:}
\bea
0&=&-\mathcal{R}+2(8-p)\square S + (8-p) (9-p)  \left(\partial S\right)^2+\frac{1}{2} \left(\partial \phi\right)^2\nonumber\\
   &&+e^{-2  R} \left(
   (7-p) (8-p) S'^2-\frac{1}{2} \phi'^2\right)+2(8-p) e^{- R} \mathcal{K} S'+\mathcal{K}^2-\mathcal{K}\cdot \mathcal{K}\nonumber\\
   &&-(7-p)\left( e^{-2   S} (8-p)-\frac{(7-p)}{2} e^{-\frac{(3-p)}{2}  \phi-2 (8-p)   S}\right).
\eea
\underline{Einstein- $\theta \theta$:}
\bea
0&=&-\frac{\mathcal{R}}{2}+\square R+(7-p)\square S +\left(\partial R\right)^2+\frac{1}{2}
   (7-p) (8-p)  \left(\partial S\right)^2+(7-p)  \left(\partial R\cdot \partial S\right) \nonumber\\
   &&+\frac{1}{4} \left(\partial \phi\right)^2+\frac{1}{4} e^{-2 R} \phi'^2+(7-p)e^{-2  R}
   \left(\frac{(8-p)}{2} S'^2+S''-S'R'\right)\\
   &&+e^{-R}\left(\mathcal{K}'+(7-p)\mathcal{K} S'\right)+\frac{1}{2}\left(\mathcal{K}\cdot \mathcal{K}+ \mathcal{K}^2\right).\nonumber\\
   &&-\frac{(7-p)}{2} \left(e^{-2 S} (6-p) +\frac{(7-p)}{2} e^{-\frac{(3-p)}{2}  \phi-2 (8-p)  S}\right).
\eea
\underline{$\phi$ field equation:}
\bea
0&=&\square \phi + \partial R\cdot \partial \phi+(8-p)   \partial S \cdot \partial \phi+e^{-2  R}
   \left(\phi''- R' \phi'+(8-p) S' \phi'\right)\nonumber\\
&&+e^{-R}\mathcal{K}\phi'+\frac{(3-p) (7-p)^2}{4} e^{-\frac{(3-p)}{2}  \phi-2 (8-p)  S} .
\eea
These `reduced' field equations have been verified by evaluating them explicitly in terms of general components for our metric ansatz (\ref{metricansatz}), and comparing to the unreduced equations (\ref{fe1},\ref{fe2},\ref{fe3}).

\bibliography{submit}{}
\bibliographystyle{apsrev}

\end{document}